\documentclass[epj]{svjour}

\usepackage[dvips]{graphicx}

\usepackage{array}
\usepackage{amssymb}
\usepackage{amsmath}
\usepackage{hhline}
\usepackage{longtable}
\usepackage{dcolumn}
\usepackage{bm}
\usepackage{subfigure}
\usepackage{epsfig}
\usepackage{latexsym,amsmath}
\usepackage{amsbsy}

\makeatletter
\renewcommand\appendix{\par
  \setcounter{section}{0}%
  \setcounter{subsection}{0}%
  \setcounter{equation}{0}%
  \renewcommand\theequation{\Alph{section}.\arabic{equation}}
  \renewcommand\thesection{Appendix \@Alph\c@section:}}
\makeatother

\begin{document}
\title{Mean first-passage time for random walks on undirected networks}

\author{Zhongzhi Zhang\inst{1,2} \thanks{e-mail: zhangzz@fudan.edu.cn} \and Alafate Julaiti\inst{1,2} \and Baoyu Hou \inst{3} \and Hongjuan Zhang \inst{3} \and Guanrong Chen \inst{4} \thanks{e-mail: eegchen@cityu.edu.hk}}

\institute{School of Computer Science, Fudan University, Shanghai
200433, China \and Shanghai Key Lab of Intelligent Information
Processing, Fudan University, Shanghai 200433, China \and Department of Mathematics, College of Science,
Shanghai University, Shanghai 200444, China \and Department of Electronic Engineering, City University of Hong Kong, Hong Kong, China}

\date{Received: date / Revised version: date}

\abstract{In this paper, by using two different techniques we derive an explicit formula for the mean first-passage time (MFPT) between any pair of nodes on a general undirected network, which is expressed in terms of eigenvalues and eigenvectors of an associated matrix similar to the transition matrix. We then apply the formula to derive a lower bound for the MFPT to arrive at a given node with the starting point chosen from the stationary distribution over the set of nodes.
We show that for a correlated scale-free network of size $N$ with a degree distribution $P(d)\sim d^{-\gamma}$, the scaling of the lower bound is $N^{1-1/\gamma}$. Also, we provide a simple derivation for an eigentime identity. Our work leads to a comprehensive understanding of recent results about random walks on complex networks, especially on scale-free networks.
} 

 \maketitle

\section{Introduction}

As a new interdisciplinary subject field~\cite{Ne11}, complex networks received considerable interest from the scientific community in the past decade~\cite{AlBa02,DoMe02,BoLaMoChHw06}. A central issue in this area is to uncover the effects of structural properties on dynamical processes on networks~\cite{Ne03,DoGoMe08}. As a paradigmatic dynamical process, random walks on complex networks have been investigated intensively~\cite{NoRi04,SoRebe05,Bobe05,CoBeTeVoKl07,GaSoHaMa07,TeBeVo09,RoHa11,TeBeVo11}. In particular, considerable attention has been paid to evaluating the mean first-passage time (MFPT) from one node to another in networks, which is required in various settings~\cite{Re01} and has been simulated by studies of related problems on complex networks such as transport~\cite{GaSoHaMa07} and search~\cite{BeCoMoSuVo05,Sh06,BeLoMoVo11}.

The MFPT $T_{ij}$ from node $i$ to node $j$ is the expected time taken by a walker to first reach node $j$ starting from node $i$. Recently, there has been increasing interest in determining the MFPT $\langle T_j \rangle$ to a given node $j$, defined as the average of $T_{ij}$ over all starting points $i$, and in finding the dependence of $\langle T_j \rangle$ on network size $N$, since this quantity is widely used to characterize transport efficiency~\cite{Mo69,KaBa02PRE,KaBa02IJBC,BeTuKo10,Ag08,HaRo08,LiWuZh10,KiCaHaAr08,ZhQiZhXiGu09,ZhGuXiQiZh09,ZhZhXiChLiGu09,ZhXiZhLiGu09,ZhGaXi10,AgBu09,ZhLiGoZhGuLi09,ZhXiZhGaGu09,ZhLiMa11}. Previous works have unveiled nontrivial scaling of MFPT $\langle T_j \rangle$ in a network with size $N$. For example, it was shown that in some scale-free networks the MFPT $\langle T_j \rangle$ to a most-connected node $j$ behaves sublinearly with the network size $N$~\cite{KiCaHaAr08,ZhQiZhXiGu09,ZhGuXiQiZh09}, which is in sharp contrast to that in standard regular fractals (e.g., Sierpinski gaskets~\cite{KaBa02PRE,KaBa02IJBC,BeTuKo10} and $T$-fractals~\cite{Ag08,HaRo08,LiWuZh10}, where $\langle T_j \rangle$ scales superlinearly with $N$. This striking finding has inspired a growing number of works with an attempt to explore the impacts of other topological features on the scaling of MFPT $\langle T_j \rangle$ to a target node $j$, including modularity~\cite{ZhLiGoZhGuLi09}, fractality~\cite{ZhXiZhGaGu09,ZhLiMa11}, and so forth. However, various topological properties are closely related to each other, e.g., scale-free networks are often simultaneously small-world~\cite{CoHa03}, and fractal characteristic is frequently accompanied by a modular structure~\cite{SoHaMa06}. It is thus very hard to distinguish which individual property plays a crucial role in determining the behavior of MFPT $\langle T_j \rangle$ in a network. In some cases, it is possible to confuse and even misinterpret the effects of different topological features.

In this paper, we present a general framework for studying discrete-time random walks on networks, which can unify previous results about MFPT. For this purpose, we derive an exact formula of MFPT between two nodes on an arbitrary network, in terms of eigenvalues and eigenvectors of a matrix similar to the transition probability matrix associated with a random walk. Based on this formula, we further provide a lower bound of MFPT to a target for a walker, which started from a node selected from the stationary distribution over all nodes in the network. We find that the lower bound is proportional to the reciprocal of the degree of the target node, regardless of other topological properties. Particularly, in correlated networks with $N$ nodes obeying a power-law degree distribution $P(d) \sim d^{-\gamma}$, the dominating behavior of the low bound of MFPT to a hub node is $N^{1-1/\gamma}$. To that end, we make a comprehensive analysis of existing results about MFPT in different networks and show that those results can be understood within this framework. Finally, as a byproduct, we derive Kemeny's constant~\cite{LeLo02} for random walks on networks.

\section{Brief introduction to random walks on networks \label{RanWalk}}

We study discrete-time unbiased random walks~\cite{NoRi04} on an arbitrary connected undirected network $G$ with $N$ nodes and $M$ edges. The $N$ nodes of $G$ are denoted by $1,2,3,\ldots, N$. The connectivity of $G$ is encoded in its adjacency matrix $A$, whose entry $a_{ij}=1$ (or 0) if nodes $i$ and $j$ are (not) connected by an edge. The degree of node $i$ is $d_i=\sum_{j=1}^{N} a_{ij}$, and the diagonal degree matrix of $G$, denoted by $D$, is defined as: the $i$th diagonal element is $d_i$, while all non-diagonal elements are zero. Thus, the total degree of all nodes is $K=2M= \sum_{j=1}^{N}d_i$, and the average degree is $\langle d \rangle=2M /N$.

In the process of a random walk considered here, at each time step the walker starting from its current location moves with a uniform probability to one of its neighboring nodes. In fact, this stochastic process of random walks is characterized by the transition matrix $P=D^{-1}A$, whose element $p_{ij}= a_{ij}/d_i$ presents the probability of jumping from $i$ to $j$ in one step. Let $P^t$ express the $t$th power of matrix $P$. Its $ij$th entry denoted by $(p^{t})_{ij} $ represents the probability for a walker, starting from node $i$, to visit node $j$ in $t$ steps. It is easy to verify that such a random walk is actually an ergodic Markov chain~\cite{KeSn76,AlFi99}, whose stationary distribution $\pi=(\pi_1, \pi_2,\ldots, \pi_N)^\top$ is a unique probabilistic vector satisfying $\pi_i=d_i/K$, $\sum_{i=1}^{N}\pi_i=1$, and $\pi^{\top}P=\pi^{\top}$, where the superscript $\top$ means transpose.

\section{MFPT between two nodes \label{model}}

A quantity of primary interest related to random walks is MFPT. It encodes a great deal of information about the random-walk dynamics~\cite{Re01}. In what follows, we use two different approaches to derive MFPT between an arbitrary pair of nodes.

\subsection{Method of spectral graph theory}

It is well known that the MFPT $T_{ij}$ from node $i$ to node $j$ can be expressed exactly in terms of the elements $z_{ij}$ of the fundamental matrix $Z$ for the corresponding Markov process, defined by
\begin{equation}\label{Mat01}
Z=(I-P+W)^{-1},
\end{equation}
where $I$ denotes the identity matrix of order $N$, and $W$ is the matrix with each row being $\pi^{\top}$. Specifically~\cite{GrSn97}
\begin{equation}\label{Mat02}
T_{ij}=\frac{z_{jj}-z_{ij}}{\pi_j}.
\end{equation}

Next, using the spectral graph theory~\cite{Ch97}, we show that $T_{ij}$ can be expressed in terms of the eigenvalues and eigenvectors of a matrix related to the random walks. Since $P$ is asymmetric except when graph $G$ is regular, we introduce the following matrix:
\begin{equation}\label{Mat03}
S=D^{-\frac{1}{2}}AD^{-\frac{1}{2}}=D^{\frac{1}{2}}PD^{-\frac{1}{2}}.
\end{equation}
Obviously, $S$ is real and symmetric and has the same set of eigenvalues as $P$. Moreover, if $\psi$ is an eigenvector of $S$ corresponding to eigenvalue $\lambda$, then $D^{-\frac{1}{2}}\psi$ is an eigenvector of $P$ associated with eigenvalue $\lambda$.

Let $\lambda_1$, $\lambda_2$, $\lambda_3$, $\cdots$, $\lambda_N$ be the $N$ eigenvalues of matrix $S$, rearranged as
$1=\lambda_1>\lambda_2 \geq \lambda_3 \geq \ldots \geq \lambda_N \geq -1$, and let $\psi_1$, $\psi_2$, $\psi_3$, $\ldots$, $\psi_N$ be the corresponding mutually orthogonal eigenvectors of unit length, where $\psi_i=(\psi_{i1},\psi_{i2},\ldots,\psi_{iN})^{\top}$. Then
\begin{equation}\label{Mat04}
\Psi^{\top} S \Psi={\rm diag}[\lambda_1, \lambda_2, \ldots, \lambda_N],
\end{equation}
where $\Psi$ is an orthogonal matrix, that is,
\begin{equation}\label{Mat05}
\Psi \Psi^{\top}=\Psi^{\top}\Psi=I.
\end{equation}
Thus, the entry $\psi_{ij}$ of $\Psi$ satisfy
\begin{equation}\label{Mat06}
\sum_{k=1}^{N}\psi_{ik}\psi_{jk}=\sum_{k=1}^{N}\psi_{k i}\psi_{kj}=\begin{cases}
1, &{\rm if} \quad i=j,\\
0, &{\rm otherwise}.
\end{cases}
\end{equation}
Equation~(\ref{Mat04}) can be rewritten in an alternative way as
\begin{equation}\label{Mat07}
S=\Psi{\rm diag}\,[\lambda_1, \lambda_2, \ldots, \lambda_N]\,\Psi^{\top},
\end{equation}
which indicates that the entry $s_{ij}$ of matrix $S$ has the following spectral representation:
\begin{equation}\label{Mat08}
s_{ij}=\sum_{k=1}^{N}\lambda_k \psi_{k i} \psi_{kj}.
\end{equation}

We proceed to express matrices $I$, $P$, and $W$ in terms of $\Psi$ and $D$. First, according to Eq.~(\ref{Mat05}), one has
\begin{equation}\label{Mat09}
I=D^{-\frac{1}{2}}\Psi{\rm diag}\,[1, 1, \ldots, 1]\,\Psi^{\top}D^{\frac{1}{2}}.
\end{equation}
On the other hand, from Eqs.~(\ref{Mat03}) and~(\ref{Mat05}) one obtains
\begin{equation}\label{Mat10}
P=D^{-\frac{1}{2}}S D^{\frac{1}{2}}=D^{-\frac{1}{2}}\Psi{\rm diag}\,[\lambda_1, \lambda_2, \ldots, \lambda_N]\,\Psi^{\top}D^{\frac{1}{2}}.
\end{equation}
Finally, it follows from the definition of the stationary probability~\cite{KeSn76,AlFi99} that
\begin{eqnarray}\label{Mat11}
W &=& \lim_{t \rightarrow \infty} P^t=\lim_{t \rightarrow \infty} \left(D^{-\frac{1}{2}}S D^{\frac{1}{2}}\right)^t \nonumber \\
&=&\lim_{t \rightarrow \infty} D^{-\frac{1}{2}}\left(\sum_{k=1}^{N}\lambda_k^t \psi_k \psi_k^{\top}\right) D^{\frac{1}{2}}\nonumber \\
&=& D^{-\frac{1}{2}} \psi_1 \psi_1^{\top}  D^{\frac{1}{2}}+\lim_{t \rightarrow \infty} D^{-\frac{1}{2}}\left(\sum_{k=2}^{N}\lambda_k^t \psi_k \psi_k^{\top}\right) D^{\frac{1}{2}}\nonumber \\
&=& D^{-\frac{1}{2}}\Psi{\rm diag}\,[1, 0, \ldots, 0]\,\Psi^{\top}D^{\frac{1}{2}},
\end{eqnarray}
where we have used the facts $\lambda_1=1$ and $|\lambda_k| < 1$ ($k \geq 2$) when $G$ is not a bipartite graph.

The above three equations lead to
\begin{equation}\label{Mat111}
Z=D^{-\frac{1}{2}}\Psi{\rm diag}\,\left[1, \frac{1}{1-\lambda_2}, \ldots, \frac{1}{1-\lambda_N}\right]\,\Psi^{\top}D^{\frac{1}{2}}.
\end{equation}
Then, rewrite the entry $z_{ij}$ as
\begin{equation}\label{Mat12}
z_{ij}=\psi_{1i} \psi_{1j}\sqrt{\frac{d_j}{d_i}}+\sum_{k=2}^{N}\left(\frac{1}{1-\lambda_k}\psi_{ki} \psi_{kj}\sqrt{\frac{d_j}{d_i}}\right).
\end{equation}

In order to express MFPT $T_{ij}$ in terms of the eigenvalues and eigenvectors of matrix $S$, all that is left is to determine the eigenvector $\psi_{1}$ of $S$ corresponding to eigenvalue $\lambda_{1}=1$. Since $S=D^{\frac{1}{2}}P D^{-\frac{1}{2}}$ and $S \psi_{1}=\psi_{1}$, one has $P(D^{-\frac{1}{2}}\psi_{1})=(D^{-\frac{1}{2}}\psi_{1})$. Since $P1=1$, one also has $D^{-\frac{1}{2}}\psi_{1}=1$, so that
\begin{eqnarray}\label{Mat13}
\psi_{1}&=&(\psi_{11},\psi_{12},\ldots,\psi_{1N})^\top =(\sqrt{\pi_1},\sqrt{\pi_2},\ldots,\sqrt{\pi_N})^\top \nonumber \\
&=&\left(\sqrt{\frac{d_1}{K}},\sqrt{\frac{d_2}{K}},\ldots,\sqrt{\frac{d_N}{K}}\right)^\top.
\end{eqnarray}
Inserting Eqs.~(\ref{Mat12}) and~(\ref{Mat13}) into Eq.~(\ref{Mat02}) and with some calculation one obtains
\begin{equation}\label{Mat14}
T_{ij}= K \sum_{k=2}^{N}\frac{1}{1-\lambda_k} \left(\frac{\psi^2_{k j}}{d_j}-\frac{\psi_{ki} \psi_{kj}}{\sqrt{d_i d_j}}\right),
\end{equation}
which will be very useful in the following derivation of MFPT $\langle T_{j} \rangle$ to a given target node $j$.

\subsection{Method of generating functions}

Notice that Eq.~(\ref{Mat14}) can also be derived by using the formalism of generating functions~\cite{Wi94}. Let $(q^t)_{ij}$ represent the probability of a walker from node $i$ to first reach node $j$ after $t$ steps. Note that $(q^t)_{ij}$ is quite different from the $(p^t)_{ij}$ studied above, since the latter represents the probability of hitting $j$ in $t$ steps given that the walker stated from node $i$. The two quantities satisfy the following relation:
\begin{equation}\label{GenFun01}
(p^{t})_{ij}=\sum_{s=0}^{t}(q^{s})_{ij}(p^{t-s})_{jj}\,.
\end{equation}
This relation can also be expressed in terms of generating functions. To show this, we define
\begin{equation}\label{GenFun02}
\tilde{Q}_{ij}(x)=\sum_{t=0}^{\infty}(q^{t})_{ij} x^{t}\,,
\end{equation}
and
\begin{equation}\label{GenFun03}
\tilde{P}_{ij}(x)=\sum_{t=0}^{\infty}(p^{t})_{ij} x^{t}\,,
\end{equation}
where $x$ is a complex number. By the Abel theorem, $\tilde{P}_{ij}(x)$ is a uniformly continuous function for $x \in [0,1]$ and $0<\tilde{P}_{ij}(x)\leq 1$, while $\tilde{Q}_{ij}(x)$ is continuous for $x \in [0,1]$ but may diverge as $x \rightarrow 1^{-}$. Equations~(\ref{GenFun02}) and~(\ref{GenFun03}) together yield a useful relation given by
\begin{equation}\label{GenFun04}
\tilde{Q}_{ij}(x)=\frac{\tilde{P}_{ij}(x)}{\tilde{P}_{jj}(x)}\,.
\end{equation}

By its definition, the MFPT $T_{ij}$ can be obtained via differentiating $\tilde{Q}_{ij}(x)$ with respect to $x$ and then setting $x=1$:
\begin{equation}\label{GenFun05}
T_{ij}=\frac{\rm d}{{\rm d} x}\tilde{Q}_{ij}(x) \bigg |_{x=1}=\sum_{t=1}^{\infty}t(q^{t})_{ij}.
\end{equation}
Thus, our goal is reduced to proving that Eq.~(\ref{GenFun05}) is equivalent to Eq.~(\ref{Mat14}).

As shown in Eq.~(\ref{Mat11}), one has
\begin{equation}\label{GenFun06}
P^t = D^{-\frac{1}{2}}\left(\sum_{k=1}^{N}\lambda^t_k \psi_k \psi_k^{\top}\right) D^{\frac{1}{2}},
\end{equation}
so
\begin{equation}\label{GenFun07}
(p^{t})_{ij}=\sum_{k=1}^{N}\lambda^t_k\psi_{ki}\psi_{kj}\sqrt{\frac{d_{j}}{d_{i}}}\,.
\end{equation}
Thus, for $|x|\leq1$, one has $|\lambda_k x|<1$ ($k\geq 2$) and
\begin{eqnarray}\label{GenFun08}
\tilde{P}_{ij}(x)&=&\sum_{t=0}^{\infty}\sum_{k=1}^{N}(\lambda_{k}x)^{t}\psi_{ki}\psi_{kj}\sqrt{\frac{d_{j}}{d_{i}}}\notag\\
&=&\sum_{k=1}^{N}\psi_{ki}\psi_{kj}\sqrt{\frac{d_{j}}{d_{i}}}+\sum_{k=1}^{N}\sum_{t=1}^{\infty}(\lambda_{k}x)^{t}\psi_{ki}\psi_{kj}\sqrt{\frac{d_{j}}{d_{i}}}\notag\\
&=&\psi_{1i}\psi_{1j}\sqrt{\frac{d_{j}}{d_{i}}}\sum_{t=1}^{\infty}x^{t}+\sum_{k=2}^{N}\left[\sum_{t=1}^{\infty}(\lambda_{k}x)^{t}\right]\psi_{ki}\psi_{kj}\sqrt{\frac{d_{j}}{d_{i}}}\notag\\
&=&\pi_{j}\frac{x}{1-x}+\sum_{k=2}^{N}\frac{\lambda_{k}x}{1-\lambda_{k}x}\psi_{ki}\psi_{kj}\sqrt{\frac{d_{j}}{d_{i}}},
\end{eqnarray}
where we have used Eq.~(\ref{Mat06}) and $\psi_{1i}\psi_{1j}\sqrt{\frac{d_{j}}{d_{i}}}= \pi_{j}$.
Note that
\begin{equation}\label{GenFun09}
\frac{\lambda_{k}x}{1-\lambda_{k}x}=\frac{1}{1-\lambda_{k}x}-\frac{1-\lambda_{k}x}{1-\lambda_{k}x},
\end{equation}
the second term, denoted by $g_2$, on the last line of Eq.~(\ref{GenFun08}) can be recast as
\begin{eqnarray}\label{GenFun10}
g_2&=&\sum_{k=2}^{N}\frac{1}{1-\lambda_{k}x}\psi_{ki}\psi_{kj}\sqrt{\frac{d_{j}}{d_{i}}}-\sum_{k=2}^{N}\frac{1-\lambda_{k}x}{1-\lambda_{k}x}\psi_{ki}\psi_{kj}\sqrt{\frac{d_{j}}{d_{i}}}\notag\\
&=&\sum_{k=2}^{N}\frac{1}{1-\lambda_{k}x}\psi_{ki}\psi_{kj}\sqrt{\frac{d_{j}}{d_{i}}}+\psi_{1i}\psi_{1j}\sqrt{\frac{d_{j}}{d_{i}}}\notag\\
&=&\sum_{k=2}^{N}\frac{1}{1-\lambda_{k}x}\psi_{ki}\psi_{kj}\sqrt{\frac{d_{j}}{d_{i}}}+\pi_{j}.
\end{eqnarray}
Therefore,
\begin{equation}\label{GenFun11}
\tilde{P}_{ij}(x)=\pi_{j}\frac{1}{1-x}+\sum_{k=2}^{N}\frac{1}{1-\lambda_{k}x}\psi_{ki}\psi_{kj}\sqrt{\frac{d_{j}}{d_{i}}}.
\end{equation}

Analogously, $\tilde{P}_{jj}(x)$ can be derived. Applying Eq.~(\ref{GenFun07}) gives
\begin{eqnarray}\label{GenFun12}
\tilde{P}_{jj}(x)&=&\sum_{t=0}^{\infty}\sum_{k=1}^{N}(\lambda_{k}x)^{t}\psi_{kj}^{2}\notag\\
&=&\sum_{k=1}^{N}(\lambda_{1}x)^{0}\psi_{kj}^{2}+\sum_{k=1}^{N}\sum_{t=1}^{\infty}(\lambda_{k}x)^{t}\psi_{kj}^{2}\notag\\
&=&\sum_{k=1}^{N}\psi_{kj}^{2}+\psi_{1j}^{2}\sum_{t=1}^{\infty}(\lambda_{1}x)^{t}+\sum_{k=2}^{N}\left[\sum_{t=1}^{\infty}(\lambda_{k}x)^{t}\right]\psi_{kj}^{2}\notag\\
&=&1+\pi_{j}\frac{x}{1-x}+\sum_{k=2}^{N}\frac{\lambda_{k}x}{1-\lambda_{k}x}\psi_{kj}^{2}\,.
\end{eqnarray}
The last term, denoted by $g_3$, in the last line of Eq.~(\ref{GenFun12}) can be further written as
\begin{eqnarray}\label{GenFun13}
g_3&=&\sum_{k=2}^{N}\frac{1}{1-\lambda_{k}x}\psi_{kj}^{2}-\sum_{k=2}^{N}\frac{1-\lambda_{k}x}{1-\lambda_{k}x}\psi_{kj}^{2}\notag\\
&=&\sum_{k=2}^{N}\frac{1}{1-\lambda_{k}x}\psi_{kj}^{2}-1+\psi_{1j}^{2}\notag\\
&=&\sum_{k=2}^{N}\frac{1}{1-\lambda_{k}x}\psi_{kj}^{2}-1+\pi_{j}\,.
\end{eqnarray}
Plugging Eq.~(\ref{GenFun13}) into  Eq.~(\ref{GenFun12}) leads to
\begin{equation}\label{GenFun14}
\tilde{P}_{jj}(x)=\pi_{j}\frac{1}{1-x}+\sum_{k=2}^{N}\frac{1}{1-\lambda_{k}x}\psi_{kj}^{2}.
\end{equation}

Substituting the above obtained expressions of $\tilde{P}_{ij}(x)$ and $\tilde{P}_{jj}(x)$ into Eq.~(\ref{GenFun04}) yields
\begin{eqnarray}\label{GenFun15}
\tilde{Q}_{ij}(x)&=&\frac{\pi_{j}\frac{1}{1-x}+\displaystyle\sum_{k=2}^{N}\frac{1}{1-\lambda_{k}x}\psi_{ki}\psi_{kj}\sqrt{\frac{d_{j}}{d_{i}}}}{\pi_{j}\frac{1}{1-x}+\displaystyle \sum_{k=2}^{N}\frac{1}{1-\lambda_{k}x}\psi_{kj}^{2}}\notag\\
&=&\frac{\pi_{j}+(1-x)\displaystyle\sum_{k=2}^{N}\frac{1}{1-\lambda_{k}x}\psi_{ki}\psi_{kj}\sqrt{\frac{d_{j}}{d_{i}}}}{\pi_{j}+(1-x)\displaystyle \sum_{k=2}^{N}\frac{1}{1-\lambda_{k}x}\psi_{kj}^{2}}\,.
\end{eqnarray}
Then the derivative of $\tilde{Q}_{ij}(x)$ with respect to variable $x$ is
\begin{figure*}
\begin{eqnarray}\label{GenFun16}
\frac{\rm d}{{\rm d} x}\tilde{Q}_{ij}(x)&=&\frac{\left[-\displaystyle\sum_{k=2}^{N}\frac{1}{1-\lambda_{k}x}\psi_{ki}\psi_{kj}\sqrt{\frac{d_{j}}{d_{i}}}+(1-x)\displaystyle\sum_{k=2}^{N}\frac{\lambda_{k}}{({1-\lambda_{k}x})^{2}}\psi_{ki}\psi_{kj}\sqrt{\frac{d_{j}}{d_{i}}}\right] \left[\pi_{j}+(1-x)\displaystyle\sum_{k=2}^{N}\frac{1}{1-\lambda_{k}x}\psi_{kj}^{2} \right]}{ \left[\pi_{j}+(1-x)\displaystyle\sum_{k=2}^{N}\frac{1}{1-\lambda_{k}x}\psi_{kj}^{2} \right]^{2}}\notag\\
&\quad&-\frac{ \left[-\displaystyle\sum_{k=2}^{N}\frac{1}{1-\lambda_{k}x}\psi_{kj}^{2}+(1-x)\displaystyle\sum_{k=2}^{N}\frac{\lambda_{k}}{(1-\lambda_{k}x)^{2}}\psi_{kj}^{2} \right] \left[\pi_{j}+(1-x)\displaystyle\sum_{k=2}^{N}\frac{1}{1-\lambda_{k}x}\psi_{ki}\psi_{kj}\sqrt{\frac{d_{j}}{d_{i}}} \right]}{ \left[\pi_{j}+(1-x)\displaystyle\sum_{k=2}^{N}\frac{1}{1-\lambda_{k}x}\psi_{kj}^{2} \right]^{2}}.
\end{eqnarray}
\end{figure*}
At $x=1$, all items with $(x-1)$ will be zero. So,
\begin{eqnarray}\label{GenFun17}
T_{ij}&=&\frac{\rm d}{{\rm d} x}\tilde{Q}_{ij}(x) \bigg |_{x=1}\notag \\
&=&\frac{-\pi_{j}\displaystyle\sum_{k=2}^{N}\frac{1}{1-\lambda_{k}}\psi_{ki}\psi_{kj}\sqrt{\frac{d_{j}}{d_{i}}}+\pi_{j}\displaystyle\sum_{k=2}^{N}\frac{1}{1-\lambda_{k}}\psi_{kj}^{2}}{\pi_{j}^{2}}\notag\\
&=&\frac{K}{d_{j}}\sum_{k=2}^{N}\frac{1}{1-\lambda_{k}}\left(\psi_{kj}^{2}-\psi_{ki}\psi_{kj}\sqrt{\frac{d_{j}}{d_{i}}}\right)\notag\\
&=& K \sum_{k=2}^{N}\frac{1}{1-\lambda_k} \left(\frac{\psi^2_{k j}}{d_j}-\frac{\psi_{ki} \psi_{kj}}{\sqrt{d_i d_j}}\right)\,,
\end{eqnarray}
which is consistent with Eq.~(\ref{Mat14}). Notice that the formula provided in Eqs.~(\ref{Mat14}) and~(\ref{GenFun17}) has been previously derived in~\cite{Lo96} by using a different approach.

\section{MFPT to a given node}

After deriving the MFPT $T_{ij}$ from node $i$ to node $j$ in a general network $G$, we now study the trapping problem defined on $G$, a special random walk with a deep trap located at an arbitrarily given node, e.g., node $j$. Let $\langle T_{j}\rangle$ be the average trapping time (ATT), which is the mean of MFPT $T_{i j}$ to the trap node $j$, taken over all starting points. There are two slightly different definitions for $\langle T_{j}\rangle$: For the first one, the average $\langle \cdot \rangle$ is taken over the uniform distribution of the set of starting points~\cite{Mo69,KaBa02PRE,KaBa02IJBC,BeTuKo10,Ag08}, while for the other, the average $\langle \cdot \rangle$ is taken over the stationary distribution~\cite{TeBeVo09}. Since both definitions lead to an identical scaling of $\langle T_{j}\rangle$ with network size $N$~\cite{TeBeVo09}, here we adopt the second one, given by
\begin{equation}\label{ATT01}
\langle T_{j}\rangle=\frac{1}{1-\pi_j}\sum_{i=1}^{N}\pi_i\,T_{ij}\,.
\end{equation}

Substituting the expression of $T_{ij}$ in Eq.~(\ref{Mat14}) to  Eq.~(\ref{ATT01}) gives
\begin{eqnarray}\label{ATT02}
\langle T_{j}\rangle &=&\frac{1}{1-\pi_j}\sum_{i=1}^{N}\pi_i\,K \sum_{k=2}^{N}\frac{1}{1-\lambda_k} \left(\frac{\psi^2_{k j}}{d_j}-\frac{\psi_{ki} \psi_{kj}}{\sqrt{d_i d_j}}\right)\notag\\
&=&\frac{1}{1-\pi_j}\sum_{i=1}^{N}\frac{d_{i}}{d_{j}}\sum_{k=2}^{N}\frac{1}{1-\lambda_{k}}\left(\psi_{kj}^{2}-\psi_{ki}\psi_{kj}\sqrt{\frac{d_{j}}{d_{i}}}\right)\notag\\
&=&\frac{1}{1-\pi_j}\sum_{k=2}^{N}\left(\frac{1}{1-\lambda_{k}}\psi_{kj}^{2}\sum_{i=1}^{N}\frac{d_{i}}{d_{j}}\right)\notag \\&\quad&-\frac{1}{1-\pi_j} \sum_{k=2}^{N}\left(\frac{1}{1-\lambda_{k}}\psi_{kj}\sqrt{\frac{K}{d_{j}}}\sum_{i=1}^{N}\psi_{ki}\sqrt{\frac{d_{i}}{K}}\right)\,.
\end{eqnarray}
It then follows from Eq.~(\ref{Mat06}) that $\sum_{i=1}^{N}\psi_{ki}\sqrt{\frac{d_{i}}{K}}=\sum_{i=1}^{N}\psi_{ki} \psi_{1i}$=0. Thus, the second term is equal to zero. So
\begin{eqnarray}\label{ATT03}
\langle T_{j}\rangle &=&\frac{1}{1-\pi_j}\sum_{k=2}^{N}\left(\frac{1}{1-\lambda_{k}}\psi_{kj}^{2}\sum_{i=1}^{N}\frac{d_{i}}{d_{j}}\right)
\notag\\
&=&\frac{1}{1-\pi_j}\frac{K}{d_j}\sum_{k=2}^{N}\frac{1}{1-\lambda_{k}}\psi_{kj}^{2}\,.
\end{eqnarray}

We next derive a lower bound for $\langle T_{j}\rangle$. By Cauthy's inequality, one has
\begin{equation}\label{ATT04}
\left(\sum_{k=2}^{N}\frac{1}{1-\lambda_{k}}\psi_{kj}^{2}\right)\left(\sum_{k=2}^{N}(1-\lambda_{k})\psi_{kj}^{2}\right)\geq \left(\sum_{k=2}^{N}\psi_{kj}^{2}\right)^2\,.
\end{equation}
By Eq.~(\ref{Mat08}), in matrix $S$, each of its diagonal entry $s_{jj}=\sum_{k=1}^{N}\lambda_k  \psi_{kj}^2\geq 0$ ($j=1,2,\ldots,N$). On the other hand, the traces of matrices $S$ and $P$ are both equal to zero, so $s_{jj}=0$. Thus, since $\lambda_{1}=1$, one has
\begin{equation}\label{ATT05}
\sum_{k=2}^{N}(1-\lambda_{k})\psi_{kj}^{2}=\sum_{k=1}^{N}(1-\lambda_{k})\psi_{kj}^{2}=1-\sum_{k=1}^{N}\lambda_{k}\psi_{kj}^{2}=1\,.
\end{equation}
In addition,
\begin{equation}\label{ATT06}
\sum_{k=2}^{N}\psi_{kj}^{2}=\sum_{k=1}^{N}\psi_{kj}^{2}-\pi_{j}=1-\pi_{j}\,.
\end{equation}
Therefore,
\begin{equation}\label{ATT071}
\sum_{k=2}^{N}\frac{1}{1-\lambda_{k}}\psi_{kj}^{2} \geq (1-\pi_j)^2
\end{equation}
and
\begin{equation}\label{ATT07}
\langle T_{j}\rangle \geq \frac{1}{1-\pi_j}\frac{K}{d_j}(1-\pi_j)^2=\frac{K}{d_j}(1-\pi_j)=\frac{K}{d_j}-1.
\end{equation}
This lower bound is sharp in the sense that it can be achieved in some graphs, e.g., a complete graph with $N$ nodes where the MFPT from an arbitrary node to node $j$ is $N-1$~\cite{Bobe05}, which is exactly equal to $\frac{K}{d_j}-1$ since in this case $K=N(N-1)$ and $d_j=N-1$.

Note that most real networks are sparse with the average degree $\langle d \rangle=\frac{K}{N}$ being a small constant. Thus, in these networks, for a target node with degree $d$, the lower bound becomes $\frac{N\langle d \rangle}{d}-1$, which, in the limit of large network size $N$, is proportional to $N$ and the inverse degree of the trap node. Moreover, it has been realized that most real-world networks display ubiquitous
degree correlations~\cite{Newman02,Newman03c}. At the same time, they are scale-free with their degree distributions $P(d)$ following a power-law form $P(d)\sim d^{-\gamma}$~\cite{AlBa02,DoMe02,BoLaMoChHw06}, implying that there exists hub nodes in these correlated networks whose degree is $d_{\rm max}\sim N^{1/(\gamma-1)}$~\cite{CoErAvHa01,DoGoMe08}. These nodes play a dominant role in various dynamics running on networks~\cite{Ne03,DoGoMe08}, including the trapping problem addressed here. Noticeably, when a trap is fixed at a hub node with degree $d_{\rm max}$, Eq.~(\ref{ATT07}) shows that the lower bound for ATT is $\frac{N\langle d \rangle}{d_{\rm max}}-1$. In large networks (i.e., $N\rightarrow \infty$) the scaling of ATT is $N^{1-1/(\gamma-1)}$, which grows sublinearly with the network size $N$.

To this end, we have presented a unifying framework to encompass many existing works about ATT on networks, especially on scale-free networks with degree correlations among nodes. First, in a generic scale-free network, although the low bound [Eq.~(\ref{ATT07})] for ATT to a hub node is hard to reach, but the minimal scaling $N^{1-1/(\gamma-1)}$ can be achieved in some particular scale-free networks, e.g., hierarchical and modular scale-free networks~\cite{AgBu09,ZhLiGoZhGuLi09} that thus have the most efficient structure for transport by diffusion when the target is positioned on a hub, even on a particular non-hub node~\cite{AgBuMa10}. Second, in some scale-free networks, although the scaling for ATT to a hub increases sublinearly with the network size, it cannot achieve the minimal scaling $N^{1-1/(\gamma-1)}$. Previously studied network examples include the pseudofractal web~\cite{ZhQiZhXiGu09}, the Apollonian network~\cite{ZhGuXiQiZh09}, and the $(1,3)-$flower~\cite{ZhXiZhLiGu09}. Third, the ATT to a hub can also display a linear scaling in some scale-free networks, such as the Koch networks~\cite{ZhZhXiChLiGu09,ZhGaXi10} and the $(2,2)-$flower~\cite{ZhXiZhLiGu09}. Finally, in a certain scale-free network family, the ATT to a hub can even scale superlinearly with the network size~\cite{ZhXiZhGaGu09,ZhLiMa11}.

\section{Alternative derivation of eigentime identity}

As can be easily seen from Eq.~(\ref{ATT03}), the MFPT $\langle T_{j}\rangle$ to a target node $j$ is strongly affected by the degree of the target node $j$. However, the average of MFPT $\langle T_{ij}\rangle$ from node $i$ to a node $j$ randomly chosen from all nodes accordingly to the stationary distribution is independent of the starting point $i$. Let $\langle \overline{T}_{i}\rangle$ denote this quantity corresponding to starting node $i$. Then
\begin{equation}\label{Cons01}
\langle \overline{T}_{i}\rangle=\sum_{j=1}^{N}\pi_j\,T_{ij}\,.
\end{equation}
It was shown in~\cite{AlFi99} that the sum $\sum_{i=1}^{N}\pi_j\,T_{ij}$ does not depend on $i$.  Using different methods~\cite{LeLo02,AlFi99,PaRe11}, it has been shown  that $\langle \overline{T}_{i}\rangle$ is a constant given by
\begin{equation}\label{Cons02}
\langle \overline{T}_{i}\rangle=\sum_{k=2}^{N}\frac{1}{1-\lambda_{k}}\,,
\end{equation}
which is referred to as an eigentime identity~\cite{AlFi99}, sometimes Kemeny's constant~\cite{LeLo02}.

Below we use an alternative yet simple way to re-derive Eq.~(\ref{Cons02}). According to Eqs.~(\ref{Mat02}) and~(\ref{Mat12}), one has
\begin{eqnarray}\label{Cons03}
\langle \overline{T}_{i}\rangle&=&\sum_{j=1}^{N}\pi_j\,T_{ij}=\sum_{j=1}^{N}(z_{jj}-z_{ij}) \notag \\
&=&\sum_{j=1}^{N}\sum_{k=2}^{N}\frac{1}{1-\lambda_k}\left(\psi_{kj}^2-\psi_{ki} \psi_{kj}\sqrt{\frac{d_j}{d_i}}\right)\notag \\
&=&\sum_{k=2}^{N}\frac{1}{1-\lambda_k}\sum_{j=1}^{N}\psi_{kj}^2\notag \\&\quad&-\sum_{k=2}^{N}\frac{1}{1-\lambda_k}\psi_{ki}\sqrt{\frac{K}{d_i}}\left(\sum_{j=1}^{N}\psi_{kj}\sqrt{\frac{d_j}{K}}\right).
\end{eqnarray}
It then follows from Eq.~(\ref{Mat06}) that $\sum_{j=1}^{N}\psi_{kj}^2=1$ and $\sum_{j=1}^{N}\psi_{kj}\sqrt{\frac{d_{j}}{K}}=\sum_{j=1}^{N}\psi_{kj} \psi_{1j}$=0. Thus,  Eq.~(\ref{Cons03}) reduces to Eq.~(\ref{Cons02}).

\section{Conclusions}

In this paper, we studied theoretically some aspects of random walks on general networks. A discrete-time random walk on a network is described by its transmission matrix, and thus information about the random walk process is encoded in some related matrices. Through applying two different approaches, we demonstrated how the MFPT between two nodes can be expressed in terms of the eigenvalues and eigenvectors of a matrix similar to the transmission matrix. Then, we derived an exact formula for MFPT to a given target node by expressing it in terms of the eigenvalues and eigenvectors of the same matrix, based on which we further provided a low bound for this interesting quantity. We showed that the low bound depends only on the degree of the target node and is seemingly independent of the structural properties of the network. Particularly, for scale-free networks with degree correlations, we expressed the minimal scaling of this low bound for a most-connected target node by the degree distribution exponent $\gamma$, which displays a sublinear dependence on the network size. This provides a broader view of previously reported results for MFPT to a given node on different networks, especially on scale-free networks.

In addition to the above-mentioned main results, some of our byproduct results are also very useful. For example, we gave a spectral representation for the entries of the fundamental matrix of the studied random walks. Making use of this expression, we provided a simple derivation for the eigentime identity. As another application, this spectral representation can be used to prove the following symmetry property about random walks on a network: $T_{ab}+ T_{bc} +T_{ca}= T_{ac}+ T_{cb} +T_{ba}$, which holds for three arbitrary nodes ($a$, $b$, and $c$) and was first discovered in~\cite{CoTeWi93}. We expect that this important result can find many other applications. Finally, we believe that this work can help better understand random walks on general networks, to provide useful insights in related dynamical processes on the networks.

\section*{Acknowledgment}

This work was supported by the National Natural Science Foundation
of China under Grant No. 61074119 and the Hong Kong Research Grants
Council under the GRF Grant CityU 1114/11E.


\end{document}